\documentclass[letterpaper,twocolumn,10pt]{article}
\usepackage{usenix,epsfig,endnotes}
\usepackage{xcolor}
\usepackage{graphicx,subcaption} 
\usepackage{multirow}
\usepackage{listings}
\definecolor{codegreen}{rgb}{0,0.6,0}
\definecolor{codegray}{rgb}{0.5,0.5,0.5}
\definecolor{codepurple}{rgb}{0.58,0,0.82}
\definecolor{backcolour}{rgb}{0.95,0.95,0.92}
\usepackage{pgfplots}
\usepackage[utf8]{inputenc}
\usepackage{cleveref}
\crefname{section}{§}{§§}
\Crefname{section}{§}{§§}
\usepackage{longtable}
\usepackage{footnote}
\makesavenoteenv{table}
\makesavenoteenv{tabular}
\usepackage{hyperref}
 
\lstdefinestyle{mystyle}{
    backgroundcolor=\color{backcolour},   
    commentstyle=\color{codegreen},
    keywordstyle=\color{magenta},
    numberstyle=\tiny\color{codegray},
    stringstyle=\color{codepurple},
    basicstyle=\footnotesize,
    breakatwhitespace=false,         
    breaklines=true,                 
    captionpos=b,                    
    keepspaces=true,                 
    numbers=left,                    
    numbersep=5pt,                  
    showspaces=false,                
    showstringspaces=false,
    showtabs=false,                  
    tabsize=2
}
 
\begin{document}

\date{}

\title{\Large \bf Stimulation and Detection of Android Repackaged Malware with Active Learning}

\author{
{\rm Aleieldin Salem}\\
Technische Universit\"{a}t M\"{u}nchen\\
Garching bei M\"{u}nchen\\
salem@in.tum.de
} 

\maketitle

\thispagestyle{empty}

\subsection*{Abstract}
Repackaging is a technique that has been increasingly adopted by authors of Android malware. The main problem facing the research community working on devising techniques to detect this breed of malware is the lack of ground truth that pinpoints the malicious segments grafted within benign apps. Without this crucial knowledge, it is difficult to train reliable classifiers able to effectively classify novel, out-of-sample repackaged malware. To circumvent this problem, we argue that reliable classifiers can be trained to detect repackaged malware, if they are allowed to request new, more accurate representations of an app's behavior. This learning technique is referred to as \emph{active learning}.

In this paper, we propose the usage of active learning to train classifiers able to cope with the ambiguous nature of repackaged malware. We implemented an architecture, {\fontfamily{cms}\selectfont Aion}, that connects the processes of stimulating and detecting repackaged malware using a feedback loop depicting active learning. Our evaluation of a sample implementation of {\fontfamily{cms}\selectfont Aion} using two malware datasets (\emph{Malgenome} and \emph{Piggybacking}) shows that active learning can outperform conventional detection techniques and, hence, has great potential to detect Android repackaged malware.

\section{Introduction}
\label{sec:intro}
Malware authors targeting Android have recently been adopting a more sophisticated technique to develop and distribute their malicious instances. In essence, they leverage the ease of decompiling, modifying the source code of, and recompiling Android applications (hereafter apps) to repackage legtimate, trusted apps with malicious payloads \cite{wei2017deep,li2017understanding,pan2017dark,tam2017evolution,zhou2012dissecting}. This breed of malware is often referred to as either \emph{piggybacked} apps \cite{li2017understanding, li2017automatically} or \emph{repackaged} malware \cite{zhou2012detecting, zhou2012dissecting}. In this paper, we use the two terms interchangeably to refer to the same type of Android malware: legitimate apps that have been grafted with malicious payloads.

Repackaged malware can be thought of as an evolution of Trojan horses. The two malware breeds differ in one key aspect viz., the originality of the app's benign functionality. On one hand, Trojan horses usually comprise benign functionalities that have been developed from scratch by the malware author for the sole purpose of presenting the app as a benign one. Furthermore, malware authors of Trojan horses seldom invest adequate amounts of time and effort in developing such fake, benign facade. The resulting benign segment is, therefore, of low quality and limited functionality (e.g., a Sudoku app with five puzzles). On the other hand, repackaged malware comprise pre-existing legitimate apps (e.g., Angry Birds), that have been amalgamated with a newly injected malicious payload (e.g., sending SMS messages to premium rate numbers \cite{fratantonio2016triggerscope}).
The primary threat that repackaged malware poses is undermining user trust in legitimate apps, their developers, and the app distribution infrastructure, which can potentially have devastating effects on the entire Android ecosystem. Unfortunately, malware authors have been increasingly adopting repackaging particularly targeting third-party marketplaces as their distribution platform. In \cite{zhou2012dissecting}, Zhou et al.\ studied 1260 Android malware instances, and concluded that more than 86\% of them were repackaged. In 2015, TrendMicro reported that 77\% of the top free 50 apps on the Google Play marketplace had fake versions, with a total of 890,482 fake apps being discovered 51\% of which were found to exhibit unwanted and/or malicious behaviors \cite{luo+2014}. More recently, Li et al.\ managed to gather piggybacked versions of around 1,400 legitimate apps \cite{li2017understanding}.

Consequently, the research community has been working towards devising methods to detect Android repackaged malware. To the best of our knowledge, the majority of such methods analyze, and perhaps execute, apps belonging to a dataset (e.g., \emph{Malgenome} \cite{zhou2012dissecting}), to extract numerical features used to train a machine learning classifier. The trained classifier is used to classify out-of-sample (hereafter test) apps as malicious and benign \cite{tam2017evolution,pan2017dark,arshad2016android}. One fundamental problem facing the research community working on detecting repackaged malware is the quality of the ground truth of the currently available datasets.
That is to say, a repackaged app is merely labeled as malicious without the knowledge of which paths depict the grafted malicious behaviors, whether there are triggers that control the execution of such behaviors, and how sophisticated those triggers are. Lacking such knowledge hinders training reliable classifiers based on the apps' dynamic behaviors, which is necessary given that malware instances have increasingly been adopting techniques (e.g., obfuscation, dynamic code loading, usage of triggers, etc.), to evade detection by conventional static-based methods \cite{wei2017deep,li2017static,rasthofer2017making,zhauniarovich2014}.

We argue that, in essence, the problem of stimulating, analyzing, and detecting Android repackaged malware is a \emph{search problem}.
During the phase of training a classifier, we are in pursuit of the segments in the training apps' source code (e.g., paths in a CFG), that depict or include the injected malicious code. Finding and executing those segments facilitates the extraction of features that resemble their runtime behavior, and enables the classifier to have a collective, accurate view of the malicious behaviors exhibited by Android repackaged malware. Similarly, to reliably classify a test app, we need to execute multiple paths to increase the probability of finding (a representation of) the injected code prior to deciding upon the app's malignance.

Needless to say, the nature and locations of the grafted malicious code in repackaged apps might vary. Given that we lack such ground truth, the training and test processes need to adopt a \emph{trial-and-error} technique.
In the training phase, for example, if a feature vector ($\hat{x_i}$) representing the runtime behavior of a training app ($a_i$) is misclassified, we assume that such feature vector represents the execution path that does not reflect the app's true nature (i.e., malicious or benign). Thus, the app ($a_i$) needs to be re-executed to explore another path that yields a feature vector ($\hat{x^{\prime}_i}$) that might help the classifier assign the app to its proper class. This process can be iterated until the maximum training accuracy is achieved, which signals the best possible representation of benign and malicious behaviors embedded in the training apps.
The number of iterations needed to achieve such maximum accuracy depicts the number of times an app needs to be executed in order to have a comprehensive overview of its behavior. For repackaged malware, it can be interpreted as the number of executions within which the malicious code is likely to execute. The feature vectors corresponding to the different executions can be used to classify the app (e.g., using majority votes), as malicious or benign. 

Within the context of machine learning, the aforementioned trial and error technique is referred to as \emph{active learning}. In this paper, we propose, implement, and evaluate an active learning architecture, called {\fontfamily{cmss}\selectfont Aion\footnote{Aion is a Greek deity associated with eternity and everlasting time. His orb, depicting repetition, resembles the behavior of our proposed active learning approach, where a feedback loop closes a circle between the processes of stimulation and detection of Android repackaged malware.}} \cite{aion2017}, to stimulate, analyze, and detect Android repackaged malware. 

Our findings show that--even with primitive stimulation and classification techniques--training a classifier using active learning improves the performance of conventional dynamic-based detection methods on test datasets and can outperform their static-based counterpars. Furthermore, the conducted experiments helped us answer questions about the number of iterations needed to train the best performing classifier, the most suitable kind of features, and the type of the best performing classifier (i.e., Random Trees, K-Nearest Neighbors, Ensemble, etc.). We believe that our experiments provide insights on how to stimulate and detect Android repackaged malware using active learning, and a benchmark against which further enhancements of the architecture (e.g., via using more sophisticated stimulation techniques) can be compared.

The contributions of this paper are:
\begin{itemize}
	\item We propose a novel architecture and approach to stimulate, analyze, and detect Android repackaged malware using active learning and investigate its applicability and performance.
	\item We implemented a framework that utilizes the proposed architecture, and made the source code, the data (i.e., gathered during evaluation), and a multitude of interactive figures plotting our results available online \cite{aion2017}.
	\item We evaluate {\fontfamily{cms}\selectfont Aion} on the recently analyzed and released \emph{Piggybacking} dataset instead of its obsolete \emph{Malgenome} counterpart, and use the former to set a classification benchmark of 72\%, similar to Zhou et al.'s 79.6\% benchmark on the latter in November 2011.
\end{itemize}

\noindent
\textbf{Organization} The rest of the paper is organized as follows. In section \ref{sec:background}, we briefly discuss fundamental concepts on which our research is built. In section \ref{sec:design} the methodology adopted during the implementation of {\fontfamily{cmss}\selectfont Aion}, its architecture, and internal structure are introduced. In section \ref{sec:evaluation}, we discuss how {\fontfamily{cmss}\selectfont Aion} was evaluated by introducing the datasets we used and the experiments conducted, and discuss the observed results in section \ref{sec:discussion}. Research efforts related to our work are enumerated in section \ref{sec:related}. Lastly, section \ref{sec:conclusion} concludes the paper.

\section{Background}
\label{sec:background}
In this section, we briefly discuss concepts fundamental to follow the remainder of the paper. We discuss our definition of repackaged malware and what it comprises, and introduce the notion of active learning and relate it to our work.
\subsection{Repackaged Malware}
\label{subsec:repackaged}
Android apps can be easily disassembled, reverse engineered, modified, and reassembled \cite{bartel2012improving}. Consequently, repackaging benign apps with malicious code segments is a straightforward process that typically involves the following activities. Firstly, a reverse engineering tool, such as {\fontfamily{lms}\selectfont Apktool} \cite{apktool2017}, is used to disassemble an app into a readable, intermediate representation (i.e., Smali). Secondly, a segment of Smali code that delivers some malicious functionality is added and merged with the original Smali code. Lastly, the app is reassembled, re-signed, and uploaded to an app marketplace, possibly with a different name.

The malicious code segments injected into benign apps usually comprise two parts viz., a malicious payload and a trigger (or a hook \cite{li2017understanding}). The former part depicts the malicious intents of the malware author (e.g., sending SMS messages to premium numbers, deleting the user's contacts, leaking the current user's GPS location, etc.). In \cite{fratantonio2016triggerscope}, Fratantonio et al.\ argue that a malicious payload need not be disconnected from the original app logic; it can distort the actual app functionality, in what they refered to as logic bombs.

The trigger is meant to transfer control to the grafted malicious payload. We can define the trigger as a set of (boolean) conditions that need to hold in order for the malicious payload to execute. For authors of repackaged malware, designing a trigger is a compromise between the likelihood of execution and stealthiness of the injected malicious payload \cite{li2017understanding}. On one hand, some malware authors may elect to maximize the likelihood of the malicious payload being executed by unconditionally calling/branching to the malicious code (e.g., deployed in a separate method or component) \cite{li2017understanding}. In this case, the trigger is a \emph{null} condition. On the other hand, some authors may give stealthiness more priority, and develop more complex triggers that activate the grafted malicious functionality depending on specific system properties (e.g., date, time, GPS location, etc.), app/system intents and notifications (e.g., android.intent.action.BOOT\_COMPLETED), custom values forwarded to the app via its authors (e.g., via SMS messages \cite{rasthofer2015current}), or app logic (e.g., $if(sum\_of\_calculation==50)$). This trend of scheduling the triggering of malicious segments has been found to be adopted by the majority of Android malware \cite{wei2017deep}.
\subsection{Active Learning}
\label{subsec:activelearning}

To the best of our knowledge, the majority of Android malware detection techniques adopt the following process. Firstly, a dataset of benign and malicious apps are gathered. Secondly, the apps in the dataset are analyzed to extract numerical features from each app. That is to say, each app is represented by a vector of features ($x_i$) and a label ($y_i$) depicting its nature (i.e., malicious or benign). The extracted feature vectors, often organized in a matrix representation ($X$), are used to train a machine learning classifier ($C$) (e.g., a support vector machine). Given a feature vector of an app ($x^*$) that has not been used during the training phase (i.e., test app), the classifier ($C$), attempts to predict its corresponding label ($y^*$) effectively classifying it as malicious or benign. To asses the prediction capability of a classifier, the original labels of the test apps, known as \emph{ground truth}, are compared to those predicted by the classifier. The percentage of correctly predicted labels is known as the classification \emph{accuracy}.

Needless to say, a number of the test apps' labels are incorrectly predicted (i.e., the corresponding app is misclassified), which negatively affects the classification accuracy. If the classification accuracy is inplausible, other classifiers can be probed, the existing features extracted from the apps are further processed (e.g., to eliminate noisy, irrelevant features), a new method of feature extraction is adopted, or a combination of those methods. This setting is referred to as \emph{passive learning} \cite{tong2001active}. The passiveness dwells in the inability of the trained classifier to ask for more accurate representations of the misclassified apps in the form of different feature vectors.

In an active learning setting, the classifier is allowed to \emph{query} the sources from whence the feature vectors have been extracted for alternative representations of the same entities \cite{tong2001active}. Within the context of Android malware detection, if a feature vector of a test app ($x^*$) is misclassified, the classifier ($C$) is allowed to instruct the feature extraction mechanism to generate another feature vectors ($x^*_{new}$) for the same app. This process can be repeated until either the misclassified app is correctly classified, or the overall classification accuracy of ($C$) has converged to a maximum value.

\section{Design and Implementation}
\label{sec:design}
\subsection{Overview}
\begin{figure}
\includegraphics[scale=0.75]{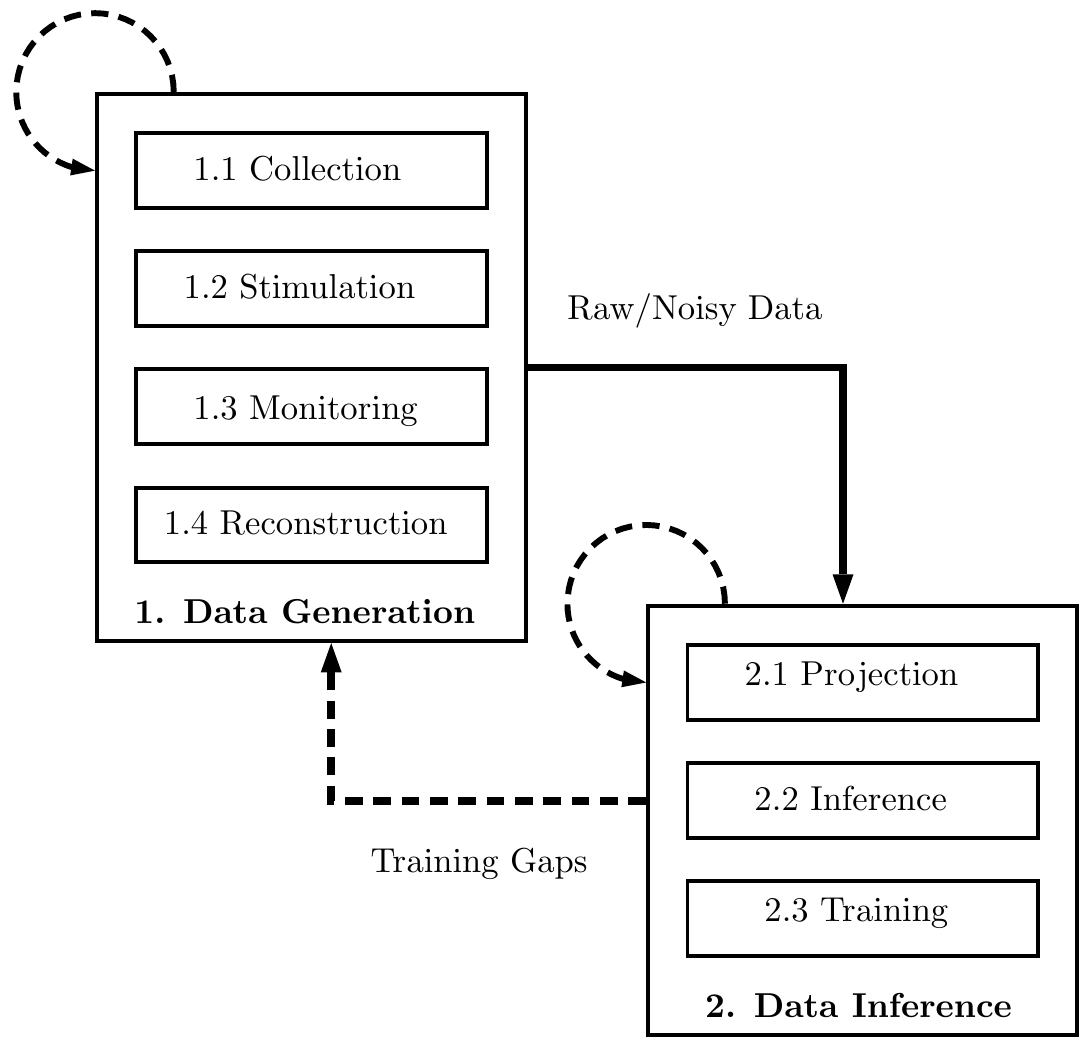}
\caption{An architecture that uses active learning to stimulate, analyze, and detect Android repackaged malware.}
\label{fig:architecture}
\end{figure}

The typical malware detection process introduced in the previous section can be grouped into two main sub processes viz., \emph{Data Generation} and \emph{Data Inference}, as seen in figure \ref{fig:architecture} which depicts our proposed architecture {\fontfamily{cmss}\selectfont Aion}.

The former process commences with the acquisition of a dataset of malicious and benign Android apps. Using a \emph{Stimulation} method, the collected apps are analyzed statically and executed within a controlled, virtualized environment to gather information about the apps' runtime behavior. Within the domain of malware analysis, there are different non-/invasive approaches to stimulating an app \cite{xue2017malton}. For example, Abraham et al.\ \cite{abraham2015} implemented an approach meant to force the execution of particular branches within an app that contain suspicious API calls, whereas Rasthofer et al.\ \cite{rasthofer2017making} utilized symbolic execution to devise environments (i.e., a set of values) required to execute branches containing specific type of API calls. The monitoring component depicted in the figure is responsible for recording the the stimulated behavior for further analysis. It can be deployed on a remote server, on the environment within which an app is executed, embedded within the app itself (e.g., as logging statements), or a combination of those techniques. Lastly, the monitored behavior may need to be further processed to represent it in a parseable, understandable format. For example, the API calls issued by an app during runtime are usually reorganized in the format of comma-separated strings (i.e., traces).

The stimulated, monitored, and reconstructed representations of the apps' behaviors are usually raw, and need to be further processed before they are used for training a classifier. Data Inference is concerned with inferring relevant information from the raw data received from its Data Generation counterpart that might facilitate segregating the two classes of apps. The \emph{Projection} component is reponsible for processing the raw data and projecting it into a different representation or dimensionality. For example, a trace of API calls can be processed to omit those calls that do not manipulate sensitive system resources (e.g., camera) or extract numerical features (e.g., counts of different API calls). To further remove noise from projected data, the \emph{Inference} component attempts to extract patterns and information that might facilitate the classification of apps. Depending on the format of the projected data, this component can either apply feature selection techniques to rule out irrelevant features, or infer rules that depict characteristics unique to a class of apps. For instance, by studying the API call traces of apps, a rule can be inferred that malicious apps usually use a particular set/sequence of API calls. Lastly, the \emph{Training} component handles training a classifier using the processed raw data, validates the results (e.g., via a test dataset), and reports classification results.

The two dashed arrows looping out of and into the two processes imply that the operations within such processes can be perfomed multiple times before the process reports its final output. For example, the Data Generation process can continue to add new apps to its repository, stimulate them multiple times, monitor their runtime behaviors, and report an augmented version of the different stimulation sessions. Similarly, the Data Inference process is allowed to consult various combinations of projection, inference, and classification techniques in pursuit of the best classification accuracy.

\begin{figure}
\centering
\includegraphics[scale=0.6]{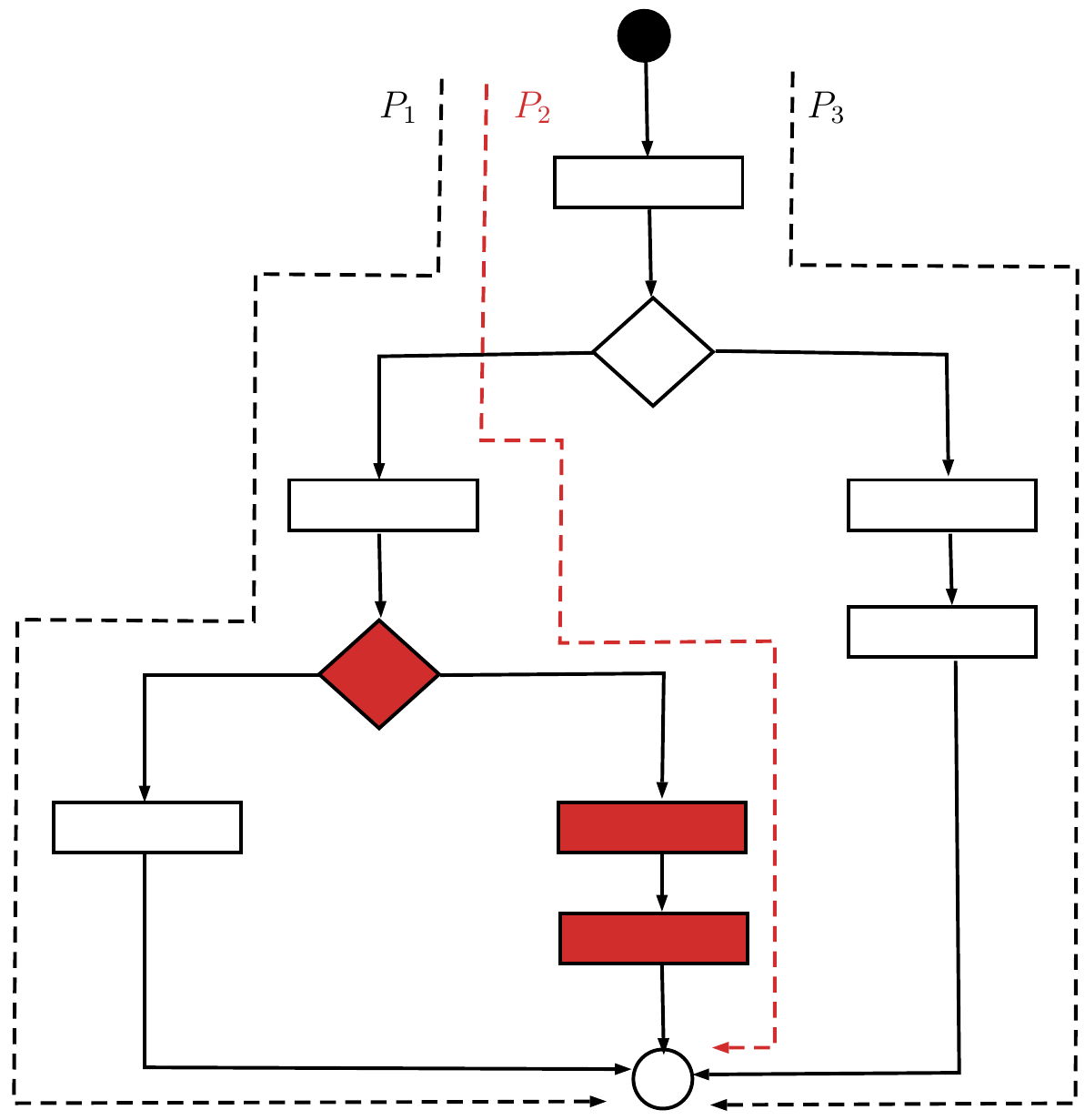}
\caption{An example of a repackaged malware's control flow graph (CFG). The colored segment depicts the malicious payload injected into the app, whereas the dashed arrows represent different paths through the CFG.}
\label{fig:cfg}
\end{figure}

Lastly, the dashed arrow extending from the Data Inference process to its Data Generation counterpart depicts the active learning aspect of our proposed approach. We illustrate the functionality and significance of such mechanism using the control flow graph (CFG) in figure \ref{fig:cfg}. Consider such CFG as that of a benign app ($a_i$) that has been injected with malicious segments (colored blocks). Regardless of the technique it employs to target and execute specific branches, the stimulation component should devise test cases that execute the path ($P_2$) which includes the malicious segments. The feature vectors extracted from the representation of ($P_2$) (e.g., API call trace) should depict the malicious behavior grafted into the benign app and, in turn, help the trained classifier assign this app to the malicious class. Nevertheless, since the location of the injected malicious payload is unknown, and since the malicious payload is assumed to be smaller than the original benign code, the stimulation component is likely to target other branches/statements within the CFG, effectively executing other paths (i.e., $P_1$ or $P_3$). Feature vectors extracted from such paths are expected to represent benign behaviors, leading to the misclassification of the app. The feedback mechanism reports misclassification of ($a_i$) to the Data Generation process, particularly the stimulation component. Using this information, the latter attempts to target different branches/statements to execute different paths within the app's CFG, and forward the reconstructed behaviors of such newly-executed paths to the Data Generation process for re-training. In theory, ($a_i$) will continue to be misclassified until ($P_2$) is executed, which is expected to help the trained classifier deem the app as malicious. Consequently, the feedback mechanism will continue to instruct the stimulation component to target different code segments until such path is executed.

Given a dataset of training apps, the feedback mechanism is meant to provide the Data Generation module with the best representation of each app that resembles its nature (i.e., malicious or benign). A classifier trained with such representations is expected to possess a comprehensive view of the malicious and benign behaviors exhibited by the apps within the training dataset.
However, the large combination of different locations of the injected malicious payloads, their contents and behaviors, and their triggers significantly impedes the process of executing multiple paths per app. In other words, it can take an unforeseeable amount of time to find the best representation of each app in the training dataset. Furthermore, achieving a perfect training accuracy is unlikely, especially since the behaviors of some apps can mimick those of apps belonging to the other class. That is to say, classification errors are inevitable. In this context, we design the feedback process to terminate once the best possible training accuracy is achieved, which is subject to two constraints. Firstly, the feedback process continues as long as the performance of the classifier is increases across two iterations or decreases by a small percentage (e.g., 1\%). Secondly, we specify a maximum number of iterations, after which the training process is terminated.

\subsection{Implementation}
\label{subsec:implementation}
The exact methods, techniques, and algorithms used by different components in the two processes of Data Generation and Data Inference can vary. In other words, the proposed architecture is completely agnostic to specific methods utilized by different components. However, the use of different stimulation approaches might affect the implementation of the Data Inference components and the feedback mechanism. In this paper, we focus on demonstrating the proposed architecture and investigating the applicability of active learning to the problem at hand. Consequently, we adopt an example in which we use basic techniques for stimulation, detection, and feedback mechanisms, and leave the utilization of more sophisticated stimulation/detection methods for future work.

The stimulation technique used in this paper is based on a non-invasive, UI-based tool, called {\fontfamily{lms}\selectfont Droidutan} \cite{droidutan2017}, which is designed to emulate the user-app interaction. The tool starts the main activity of an app, retrieves its UI elements, chooses a random element out of the retrieved ones, and interactes with it. The interaction hinges on the class of the chosen UI element. For example, if the UI element is a \emph{Button}, {\fontfamily{lms}\selectfont Droidutan} will tap it, if it is a \emph{TextField}, a random text will be typed into it, and so forth. The tool replicates the same behavior with any activity that may start during the stimulation period. To simulate the occurrence of system notifications or inter-app communication, the tool randomly broadcasts intents declared and registered to by the app in its manifest file.

Interacting with an app generates a runtime behavior that we define in terms of API calls.
We use {\fontfamily{lms}\selectfont droidmon} as our monitoring component to intercept and record a particular set of API calls that interact with system resources such as the device's camera, the user's contacts, the package installer, the GPS module, etc. 
The tool records the intercepted calls along with their arguments and return results, and writes them to the system log. 
For each app, we retrieve the API calls from the log and represent the app's behavior as a series of calls, each having the format $[package.module.class].[method]$.

The APK archives of the apps under test and their corresponding traces retrieved from {\fontfamily{lms}\selectfont droidmon}'s output are used to extract static and dynamic features, respectively.
After surveying the literature, we concluded that static features can be categorized into basic metadata features extracted from the AndroidManifest.xml file \cite{sato2013detecting}, features related to the permissions required by the app \cite{sanz2013puma, huang2013performance}, and features related to the API calls found within the app's code \cite{wu2012droidmat}.
For each app, we extract static features depicting the aforementioned three categories. We refer to those features as \emph{basic}, \emph{permission}, and \emph{api} features, and we list them in appendix \ref{app:appendixa}.
The dynamic features extracted from each app's API calls trace is a feature vector of 37 attributes, each of which depicts the number of times a method belonging to one of the API packages hooked by {\fontfamily{lms}\selectfont droidmon}, listed in appendix \ref{app:appendixb}, has been encountered in the trace.

The static and dynamic feature vectors extracted from the apps' APK's and their traces are used to train a majority vote classifier (i.e., Ensemble classifier), using a total of 12 classifiers. The classifiers used to train the Ensemble classifier are K-Nearest Neighbors (KNN) with values of $K=\lbrace 10,25,50,100,250,500\rbrace$, random forests with values of $Estimators\ (E)=\lbrace 10,25,50,75,100\rbrace$, and a Support Vector Machine (SVM) with the linear kernel.

We assess the performance of different classifiers according to two metrics viz., F1 score and specificity. The first metric is defined as $F1=2\times \frac{precision\times recall}{precision+recall}$ such that precision $= \frac{TP}{TP+FP}$ and recall $=\frac{TP}{TP+FN}$. True positives (TP) denote malicious apps correctly classified as malicious, false positives (FP) denote benign apps mistakenly classified as malicious, and false negatives (FN) denote malicious apps mistakenly classified as benign. Thus, the F1 score is meant to assess the ability of a classifier to recognize malicious apps and correctly classify them. We also keep track of the classifier's performance on benign apps (i.e., whether it classifies them correctly as benign), via specificity $=\frac{TN}{TN+FP}$ where (TN) stands for true negatives: benign apps correctly classified as benign. We keep track of these two metrics to assess whether a classifier is biased towards one class in favor of the other. For example, if a classifier scores high F1 and low specificity scores, this signals its bias towards classifying the majority of apps as malicious. Detecting a classifier bias can be used to amend the methods adopted to train a classifier (e.g., the type of extracted features).

The apps misclassified by the Ensemble classifier should be, according to our proposed approach, re-stimulated to execute different segments within their code yielding different paths, different traces of API calls and, consquently, different feature vectors.
Given that {\fontfamily{lms}\selectfont Droidutan} is a random-based testing tool, each run of an app is likely to execute a different path.
Consequently, the feedback mechanism in this implementation of {\fontfamily{cms}\selectfont Aion} comprises merely re-running an app.
To ensure that random stimulation results into the execution of different paths, we kept track of the API traces recorded by {\fontfamily{lms}\selectfont droidmon} for each app.
We averaged the number of API calls that change in an apps trace across two consecutive stimulations, and found out that an average of 60 API calls (disregarding their arguments and return values) change with every stimulation.

\section{Evaluation}
\label{sec:evaluation}
In this section, we evaluate the proposed active learning based approach using two datasets of real world Android malware. The main hypothesis of this paper is that active learning enhances the performance of classifiers trained to effectively detect Android repackaged malware. Consequently, we aspire to answer the following research questions:

\noindent \textbf{Q1:} How effective and efficient are the classifiers trained using active learning in comparison to conventional analysis and detection techniques?\\
\noindent \textbf{Q2:} How reliable are the results achieved by such classifiers?\\
\noindent \textbf{Q3:} How well do active learning classifiers generalize on test datasets?\\
\noindent \textbf{Q4:} What is the type of features (e.g., static versus dynamic) and classifiers that yield the best detection accuracies on test datasets?\\

\subsection{Datasets}
We used two datasets of malicious and benign Android apps to evaluate our approach. The malicious and benign apps of the first dataset were gathered separately. The malicious apps of such dataset belong to the \emph{Malgenome} dataset \cite{zhou2012dissecting}. Malgenome originally comprised of more than 1200 malicious apps, almost 86\% of which were found to be repackaged malware instances. Prior to being discontinued in 2015, Malgenome was considered the de facto dataset of Android repackaged malware. Fortunately, the repackaged malware belonging to this dataset continue to exist within the \emph{Drebin} dataset \cite{arp2014drebin}, from whence we acquired them. To complement the first dataset with benign apps, we randomly selected and downloaded 1800 apps from Google Play store in December 2016. Consequently, we refer to this dataset as \emph{Malgenome+GPlay} in the following sections.

The second dataset, \emph{Piggybacking} \cite{li2017understanding}, is more recent and comprises around 1300 pairs of original apps along with their repackaged versions. The process of gathering the apps, labeling them as malicious and benign, and matching apps to their repackaged counterparts has been carried out between 2014 and 2017. The dataset we used contains 1355 original, benign apps and 1399 repackaged, malicious apps. The reason behind such imbalance is that some original apps have more than one repackaged version.

Lastly, the two datasets were used separately during evaluation. Consequently, some of the experiments we conducted have two variants (i.e., one for each dataset).
\subsection{Experiments}
To answer the aforementioned research questions, we designed two sets of experiments. The first set of experiments is meant to decide upon the datasets to use in evaluating the proposed active learning approach, and to set a performance benchmark against which the proposed approach is to be compared. The benchmark is set using results achieved from conventional analysis and detection methods that have been previously utilized to detect Android (repackaged) malware. We refer to this set of experiments as preliminary experiments. The second set of experiments evaluates the performance of our proposed active learning approach.

\begin{figure}
\centering
\includegraphics[scale=0.75]{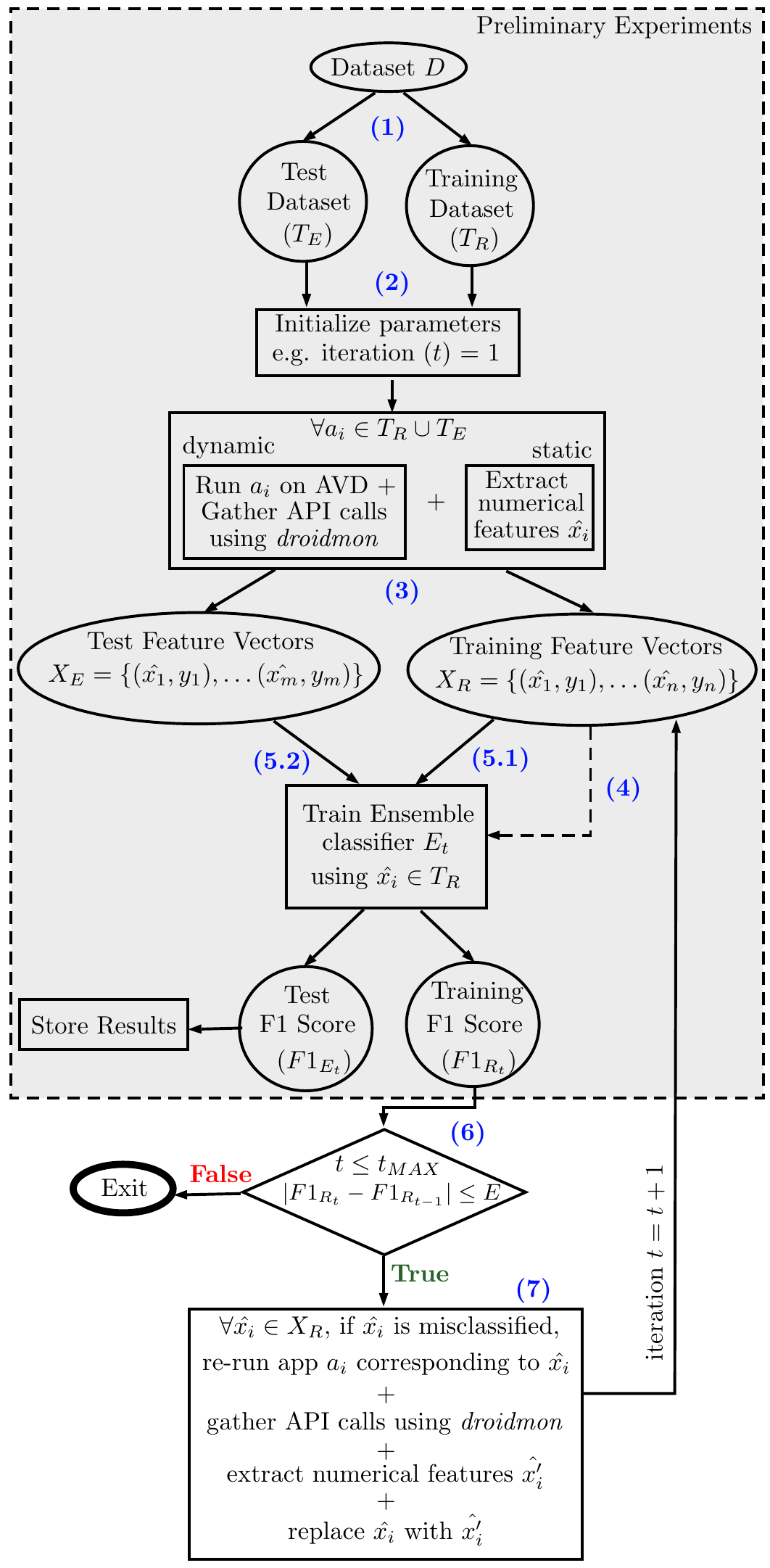}
\caption{The workflow of our experiments. Ellipses depict inputs and outputs, rectangles depict operations performed on such inputs and outputs, and rhombi depict boolean decisions. The edges between different nodes indicate the directionality of data flow. Lastly, the blue numbers are used as references to different stages of the experiment. Preliminary experiments, whether static or dynamic, conclude after stage (5) and are, hence, highlighted using the dashed gray box.}
\label{fig:experiment}
\end{figure}

Both experiment sets share a common workflow, seen in figure \ref{fig:experiment}. All experiments commence with randomly splitting a dataset of APK's into a training dataset ($T_R$) comprising two thirds of the total number of APK's and a test dataset ($T_E$) comprising the remaining one third. The second stage initializes different variables including an initial F1 score of $-1$ and a variable to keep track of the current iteration's number (i.e., $t$). The third stage is responsible for extracting a vector of numerical features ($\hat{x_i}$) for each APK ($a_i$) in both the training and test datasets ($T_R\cup T_E$). The method used to extract such numerical features hinges on the type of the experiment run. That is to say, if the experiment is \emph{dynamic}, then the APK's will be installed on an Android Virtual Device (AVD), run using {\fontfamily{lms}\selectfont Droidutan}, and monitored using {\fontfamily{lms}\selectfont droidmon}. However, if the experiment is \emph{static}, the APK's will be statically analyzed using {\fontfamily{lms}\selectfont androguard}. Note that for dynamic experiments, {\fontfamily{lms}\selectfont droidmon} may fail to produce API traces for an unforeseeable number of APK's due to either (a) the app crashing during runtime, or (b) the app not using any of the API calls monitored by the tool. The third phase concludes with producing feature vectors of numerical features for the APK's in ($T_R$) and ($T_E$). The feature vectors are stored in two datasets ($|X_R|\leq |T_R|$) and ($|X_E|\leq |T_E|$). Stage four uses the feature vectors in ($X_R$) to train a majority vote classifier ($E_t$: Ensemble classifier trained at iteration $t$), as discussed in section \ref{subsec:implementation}. In stage five, the trained classifier is validated using the vectors from ($X_R$) and tested using the vectors from ($X_E$) yielding corresponding training F1 score ($F1_{R_t}$) and test F1 score ($F1_{E_t}$).

Preliminary experiments conclude after stage five, storing the results into a SQLite database for future study. Active learning experiments compare the training accuracy ($F1_{R_t}$) scored at the current iteration ($t$) to that scored at the previous iteration ($F1_{R_{t-1}}$).
Test accuracies are not used in such comparison to prevent the test dataset (i.e., containing out-of-sample apps), from affecting the training process.
This comparison always evaluates to true after the first iteration, especially since the training F1 score is initialized to $-1$. For values of ($t> 1$), if the absolute difference between ($F1_{R_t}$) and ($F1_{R_{t-1}}$) is less than or equal to a threshold percentage ($E$) (default is $1\%$), the misclassified feature vectors will be re-run in a manner similar to stage three. Intuitively, we continue running the experiment as long as the current training F1 score ($F1_{R_t}$) is increasing. In other words, there is a potential for achieving better training, and perhaps test, scores. Furthermore, in order to accommodate for fluctuations in training accuracies, we allow the current training F1 score ($F1_{R_t}$) to drop for a maximum of ($E$). This means, however, that the experiment could go on for a prolonged period of time. To avoid this behavior, we set an upper bound ($t_{MAX}$) for the number of iterations the training phase is allowed to run. In evaluating our approach, we use the values of $10$ and $1\%$ for ($t_{MAX}$) and ($E$), respectively.

Regardless of the experiment set, a dataset is randomly split into training and test datasets at the beginning of each experiment. The training and test datasets persist throughout the experiment (i.e., through all iterations). Consequently, we run all experiments at least 25 times to increase the likelihood of different segments of the datasets being used as, both, training and test samples. Throughout implementing and evaluating {\fontfamily{cms}\selectfont Aion}, we generated a multitude of figures, which we could not include in this paper. We host and make available all our (interactive) figures, data, and source code on \cite{aion2017}.

\subsubsection{Preliminary Static Experiments}
\label{sec:exp_preliminary_static}

\begin{figure*}
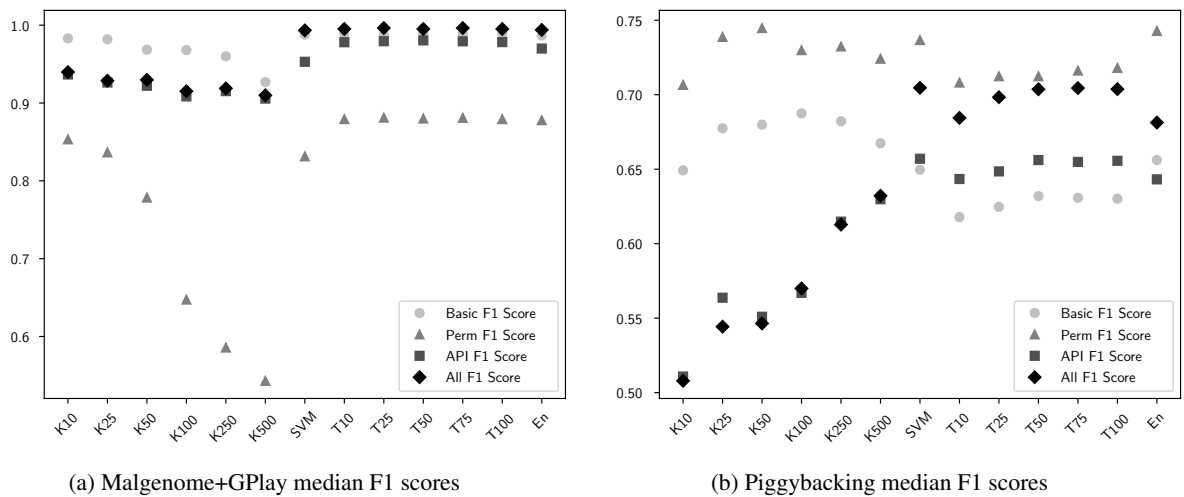

\centering
\begin{subfigure}{0.49\textwidth}
	\scalebox{.55}{\input{figs/exp_preliminary_static/Malgenome_static_F1Score_TEST_scatter.pgf}}
	\caption{Malgenome+GPlay median F1 scores}
	\label{fig:exp_pre_static_malgenomef1}
\end{subfigure}
\begin{subfigure}{0.49\textwidth}
	\scalebox{.55}{\input{figs/exp_preliminary_static/Piggybacking_static_F1Score_TEST_scatter.pgf}}
	\caption{Piggybacking median F1 scores}
	\label{fig:exp_pre_static_piggybackingf1}
\end{subfigure}
\caption{The median (after 25 runs) F1 scores (Y-axis) recorded by 13 classifiers (X-axis) using static features on the test datasets. The specificity scores are plotted in appendix \ref{app:appendixc}.}
\label{fig:exp_pre_static_f1}
\end{figure*}

In this set of experiments, only the three categories of static features, introduced in section \ref{subsec:implementation}, are used to classify apps as malicious and benign. That is to say, each app is represented by three feature vectors depicting such categories. Moreover, we combine the three feature vectors to come up with a fourth category of static features that we refer to as \emph{All}.

The scores in figure \ref{fig:exp_pre_static_f1} depict the median F1 scores recorded by different classifiers using all four categories of static features on the Malgenome+GPlay dataset and its more recent counterpart Piggybacking. To preserve space for more thorough discussion, we opted to move the specificity scores to appendix \ref{app:appendixc}.
We speculated that solely using static features to classify repackaged malware would yield mediocre classification accuracies, especially since repackaged malware usually poses as benign apps.
Nevertheless, as seen in figure \ref{fig:exp_pre_static_malgenomef1}, most classifiers perform well on the Malgenome+Play using static features, apart from permission-based features, which apparently are not informative enough to separate malicious and benign apps in this dataset.
We noticed, however, that the classifiers comparably underperform in terms of, both, the F1 and specificity scores upon applying the same method to classify apps in the Piggybacking dataset.

We believe there are two reasons behind such noticeable difference. Firstly, the malicious apps in our hybrid Malgenome+GPlay dataset have been gathered between August 2010 and October 2011 \cite{zhou2012dissecting}, whereas their benign counterparts were gathered in December 2016. Given that the Android ecosystem is continuously evolving, any two apps developed 5 years apart might look very different courtesy of newly-introduced technologies, programming interfaces, or even hardware capabilities. In fact, we noticed that there is a large difference between the sizes of APK archives belonging to apps in the Malgenome dataset (average size: 1.25MB) and those belonging to the benign apps we gathered from Google Play (average size: 16.35MB).
This difference implies, we believe, the existence of more components (e.g., activities and their classes) in recent Android apps.
Static features are built on counts and ratios of those components, the permissions they request, and the API calls they issue. Needless to say, the feature vectors extracted from apps developed in 2011 would look entirely different from those extracted from apps developed in 2015 or 2016, effectively facilitating segregating them in two classes which happen to be malicious and benign.

Secondly, despite being defined as repackaged malware, the malicious apps in Malgenome do not comply with the more recent definition of repackaged/piggybacked apps. In other words, we believe that the 86\% of apps comprising repackaged malware in the Malgenome dataset are, in fact, closer to being classic Trojan horses than being repackaged malware. The malicious apps in the Piggybacking dataset, however, comply with the definition of repackaged malware (i.e., standalone benign apps grafted with malicious payloads \cite{li2017understanding}). To empirically verify this claim, we used a compiler fingerprinting tool, {\fontfamily{lms}\selectfont APKiD}, to identify the compilers used to compile the apps in both datasets Malgenome+GPlay and Piggybacking along with a dataset of malicious apps called \emph{Drebin} \cite{arp2014drebin}. The distribution of compilers utilized by apps in such datasets is tabulated in table \ref{tab:apkid_compilers}. Using the list of compilers, we rely on the following hypothesis, originally made in \cite{rednaga2016}: if the compiler used to compile an APK is not the standard Android SDK compiler (dx) but rather a compiler used by reverse engineering tools (e.g., {\fontfamily{lms}\selectfont Apktool}) such as (dexlib), then an app is probably repackaged by a party other than the legitimate developers in possession of the app's original source code.

\begin{table}
\centering
\caption{Percentages of compilers used to compile APK's in different malicious/benign datasets as fingerprinted by {\fontfamily{lms}\selectfont APKiD}. The \emph{dexmerge} compiler is used by IDE's after \emph{dx} for incremental builds.}
\label{tab:apkid_compilers}
\begin{small}
\begin{tabular}{ccccc}
\hline
\textbf{Dataset} & \textbf{dx} & \textbf{dexmerge} & \textbf{dexlib 1.X/2.X} & \textbf{Total} \\ \hline
Drebin & 84\% & -- & 16\% & 4326 \\ \hline
Malgenome & 52\% & -- & 48\% & 1234 \\ \hline
Google Play & 61\% & 34\% & 5\% & 1882 \\ \hline
\begin{tabular}[c]{@{}c@{}}Piggybacking\\ (malicious)\end{tabular} & 22\% & 6\% & 72\% & 1399 \\ \hline
\begin{tabular}[c]{@{}c@{}}Piggybacking\\ (original)\end{tabular} & 61\% & 22\% & 17\% & 1355 \\ \hline
\end{tabular}
\end{small}
\end{table}

As seen in table \ref{tab:apkid_compilers}, the majority benign apps gathered from Google Play and the original apps in the Piggybacking dataset have been compiled using a compiler usually used within IDE's. This implies that the developers were likely in possession of the apps' source code. The same applies to the Drebin dataset which mainly comprises of malicious apps developed from scratch. However, the datasets that presumably comprise repackaged malware indicate a major difference. More than a half of the Malgenome dataset comprises apps that--according to the definitions in sections \ref{sec:intro} and \ref{sec:background}--are Trojans, whereas the majority of malicious apps in the Piggybacking dataset comply with the notion of a benign app being grafted with malicious payload and recompiled using non-standard compilers.

Such findings led us to deem the Malgenome dataset as obsolete and inaccurate.
In fact, during their analysis of a dataset of 24,650 Android malware samples, Wei et al.\ also deemed the Malgenome dataset outdated and not representative of the current Android malware landscape \cite{wei2017deep}.
Thus, devising methods to analyze and detect malicious apps in the Malgenome dataset has little benefit to the community.
Consequently, we opted to conduct the remainder of our experiments on the Piggybacking dataset, and only consider the static scores recorded on such dataset as our benchmark.

\subsubsection{Preliminary Dynamic Experiments}
\label{sec:exp_preliminary_dynamic}
The preliminary dynamic experiments are meant to resemble conventional dynamic approaches to stimulating and detecting Android repackaged malware. That is to say, apps are stimulated once prior to being processed for feature extraction and classification. In essence, such experiments depict the first iteration of our active learning experiments (i.e., without the feedback loop). The features used during such experiments are the dynamic features introduced in section \ref{subsec:implementation} along with a combination of the \emph{All} static and dynamic features that we refer to as \emph{hybrid features}.

\begin{figure}
\centering
\scalebox{.55}{\input{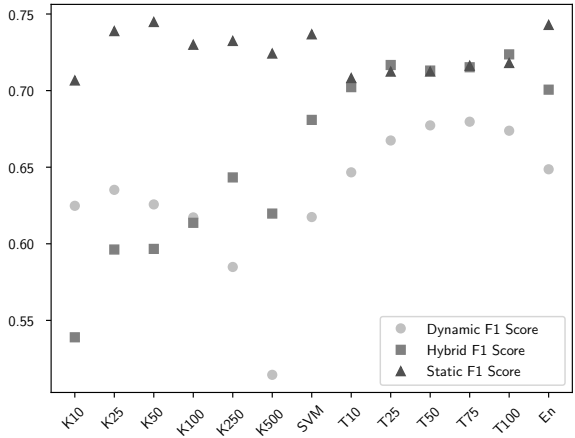}}
\caption{The median (after 25 runs) F1 scores (Y-axis) recorded by 13 classifiers (X-axis) using dynamic and hybrid features versus the best scoring static features (permission-based) on the test datasets of Piggybacking. The specificity scores are plotted in appendix \ref{app:appendixc}.}
\label{fig:exp_pre_dynamic_f1}
\end{figure}

Figure \ref{fig:exp_pre_dynamic_f1} depicts the median F1 scores achieved by different classifiers on the Piggybacking dataset after 25 runs using dynamic and hybrid features. As a reference to the performance of static features, we also plot the scores achieved by such classifiers using permission-based static features, which achieved the highest scores on the Piggbacking dataset. In figure \ref{fig:exp_pre_dynamic_f1}, dynamic features appear to be incapable of matching the detection ability of static features. One can also speculate that combining dynamic features with static features (i.e., hybrid features) decreases the latter's detection capability. Nevertheless, dynamic features perform better or slightly worse than other features in correctly classifying benign apps. To conclude, the dynamic preliminary experiments imply that dynamic features extracted after running the apps only once are initially biased towards classifying all apps as benign.
We argue that stimulating a repackaged malware once is unlikely to execute the paths under which the grafted malicious payload resides, ultimately leading to the misclassification of such app. In the next set of experiments, we attempt to verify whether re-stimulating the misclassified apps yields different paths that enhance the performance of the detection module and its classifiers.

\subsubsection{Active Learning Experiments}
\label{sec:exp_active}

Using the Piggybacking dataset, we ran at total of 25 active learning experiments, during which 4 AVD's were simultaneously used to stimulate apps using {\fontfamily{lms}\selectfont Droidutan} each for 60 seconds. The misclassified apps were allowed to be re-stimulated until either the difference in the F1 score of the Ensemble classifier drops for more than a threshold of 1\% between two iterations, or an upper bound of re-stimulations is reached. We experimented with such upper bound, and found that with 10 iterations, an experiment takes on average 26 hours to complete. Given that increasing such number did not have a noticeable effect on the achieved scores, and that it substantially increased the time taken to complete one run of the experiment, we adopted an upper bound of 10 iterations.

\begin{figure*}
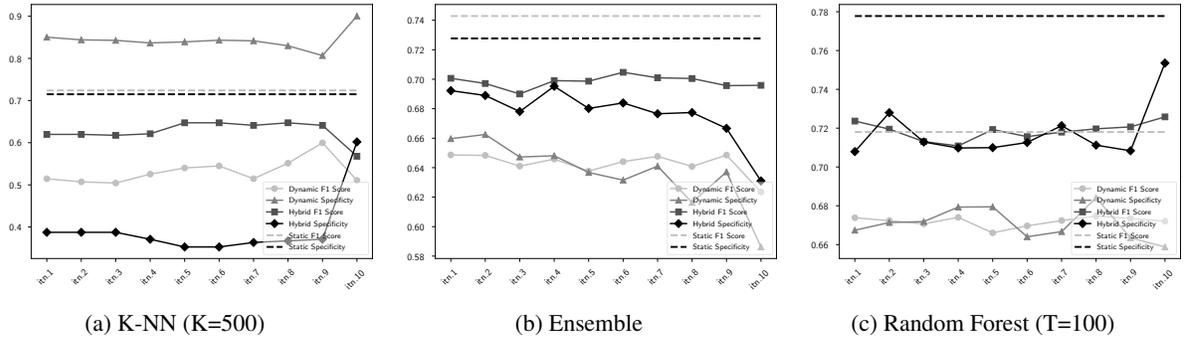

\centering
\begin{subfigure}{0.32\textwidth}
	\scalebox{.36}{\input{figs/exp_active/Piggybacking_active_f1spec_KNN500_TEST_line.pgf}}
	\caption{K-NN (K=500)}
	\label{fig:exp_active_knn500}
\end{subfigure}
\begin{subfigure}{0.32\textwidth}
	\scalebox{.36}{\input{figs/exp_active/Piggybacking_active_f1spec_Ensemble_TEST_line.pgf}}
	\caption{Ensemble}
	\label{fig:exp_active_ensemble}
\end{subfigure}
\begin{subfigure}{0.32\textwidth}
	\scalebox{.36}{\input{figs/exp_active/Piggybacking_active_f1spec_Trees100_TEST_line.pgf}}
	\caption{Random Forest (T=100)}
	\label{fig:exp_active_trees100}
\end{subfigure}
\caption{The median (after 25 runs) F1 and specificity scores (Y-axis) recorded by 13 classifiers (X-axis) for each iteration using dynamic and hybrid features versus the best scoring static features (permission-based) on the test datasets of Piggybacking.}
\label{fig:exp_active}
\end{figure*}

The scores of the lowest scoring classifier, 500-Nearest Neighbors, the highest scoring classifier, Random forest of 100 trees, and the Ensemble classifier are plotted in figure \ref{fig:exp_active}. As seen in the preliminary dynamic experiments, the figures depict the median F1 and specificity scores achieved using dynamic, hybrid, and static permission-based features achieved by the classifiers at every iteration. The static scores are plotted as straight lines, because static experiments did not comprise any iterations. Thus, the scores achieved by a classifier persist across all iterations.

We made the following observations from the plotted figures. Firstly, regardless of the used classifier, dynamic features fail to match or overperform their static counterparts. However, combining static and dynamic features boosts the performance of the latter by around 7\%. In other words, dynamic features are incapable of matching static features on their own.

Secondly, the iteration on which the maximum F1 score is achieved depends on the utilized classifier and features type. Furthermore, both the F1 and specificity scores fluctuate across iterations, and do not maintain any specific ascending or descending pattern. 

Lastly, regardless of the utilized feature type, some classifiers maintain more stable performance on both the benign and malicious apps than others. For example, for the majority of iterations, the difference between the F1 and specificity scores achieved by the K-Nearest Neighbors classifier in figure \ref{fig:exp_active_knn500} is around 20\% and 25\% for the dynamic and hybrid features, respectively. This behavior indicates bias towards one class in favor of the other. The Ensemble and Random Forest classifiers maintain more balanced performances on malicious and benign apps, which manifests in closer scores across different iterations.
\section{Discussion}
\label{sec:discussion}
In this section, we attempt to draw conclusions from the conducted experiments to answer the research questions posed earlier. 

With regard to research question \textbf{Q1}, we define the effectiveness of an active learning classifier in terms of the F1 and specificity scores it achieves using dynamic and hybrid features on the training datasets in comparison to the scores achieved by the same classifier during the static and dynamic preliminary experiments. We also consider the classifier's ability to avoid bias towards a specific class as a measure of its effectiveness. We noticed that the aforementioned scores differ from one classifier to another. For example, using dynamic features, the active learning trained 500-Nearest Neighbor classifier scored F1 and specificty scores significantly lower than what it scored during the static preliminary experiments. Furthermore, it was biased towards classifying training apps as benign, which is noticeable via its specificity score in comparison to the F1 score. However, the Ensemble and Random Forest classifiers managed to record F1 and specificity scores that are (a) greater than the scores they recorded during the static and dynamic preliminary experiments, and (b) show low bias given how close both scores are. The scores achieved by all classifiers during different types of experiments can be found on {\fontfamily{cms}\selectfont Aion}'s \href{https://aionexperiments.github.io/figures.html}{\emph{website}} \cite{aion2017}.

Similarly, the efficiency of active learning classifiers differed from one classifier to another. We define efficiency in terms of the number of iterations it takes a classifier to record its highest F1 and specificity scores on the training dataset. The 500-Nearest neighbors classifier, for instance, achieved its highest scores at the ninth iteration, whereas the Ensemble and Random Forest (with 100 Trees) classifiers recorded their highest scores at the second and third iterations, respectively. We averaged this number across all classifiers, and noticed that it takes around five iterations for an active learning classifier to achieve its maximum training accuracy. As discussed earlier, we consider this number as the number of iterations required by a classifier to have a comprehensive view of the malicious and benign behaviors within the training dataset. 

The reliability of the training scores achieved by active learning classifiers, which is the concern of \textbf{Q2}, refers to the quality of the training data used to train such classifiers. In other words, we want to make sure that the classifiers are being trained using traces and features that reflect the malicious and benign behaviors exhibited by current Android repackaged malware and apps, respectively. Otherwise, the classifiers will be trained using inaccurate data that might yield misleading results.

To illustrate this concern, consider the CFG in figure \ref{fig:cfg}. We discussed that, ideally, a repackaged malware ($a_i$) should be classified as malicious if and only if the malicious path ($P_2$) was executed. With this research question, we reason about whether the high F1 scores achieved by classifiers such as Random Trees during training honor this condition. In other words, how likely is it that a classifier deems the app ($a_i$) as benign, given that a path other than ($P_2$) has executed. Unfortunately, as discussed earlier, we do not possess any ground truth about the paths that reveal the malicious behaviors injected within apps labeled as repackaged malware. Consequently, we can only speculate about the reliability of the results. However, we reasoned about the scenarios under which this behavior could be observed (i.e., $a_i$ is classified as malicious based on benign paths, such as $P_1$).
Firstly, and presumably less likely, the original labels of the apps in the dataset may turn out to be wrong (i.e., $a_i$ is a benign or a harmless app being mistakenly labeled as malicious).
Secondly, ($P_1$) could in fact turn out to be a malicious behavior that we were not aware of and assert as benign.
Thirdly, within ($a_i$), the path ($P_1$) could be benign; however, it may have been utilized within a malicious context in other apps in the training dataset, which obliges a classifier to deem it as malicious.
In the fourth scenario, the path ($P_1$) could be a benign path that is rarely used by benign apps (i.e., anomalous), which encourages the classifier to deem it as malicious as well.
Lastly, there is the possibility of utilizing poor classifiers, features, or learning processes, which yields mediocre, unreliable classification results.

Considering an extreme scenario, assume that all repackaged malware instances in a training dataset have been correctly classified despite the fact that none of their malicious paths has executed. For this condition to hold, we need to assume that all the benign paths deemed malicious by a classifier are anomalous (i.e., seldom encountered in benign apps), which is highly unlikely given the limited number of methods a task (e.g., sending a text message), can be achieved. Moreover, if a classifier mistakenly deems the majority of benign paths as malicious, this behavior should be traceable via the specificity scores which should plumet to minimum values.

To empirically complement the assumptions above, we manually went through the repackaged malware test traces generated by {\fontfamily{lms}\selectfont droidmon} during the $25^{th}$ run of the active learning experiments. Around $75\%$ of such test traces were correctly classified using the Random Forest (T=100) classifier trained using active learning. During our analysis, we found evidence that the majority of behaviors based on which the apps were classified as malicious are, in fact, intrinsically malicious. For example, we found that a repackaged version of a simulation gaming app (i.e., \emph{com.happylabs.happymall}), was classified as malicious based on a trace that included retrieving the device's IMEI, IMSI, and network operator. Furthermore, such trace included the decryption of a URL that pointed to \emph{http://csapi.adfeiwo.com:9999}. Needless to say, such behavior is not expected from a gaming app.

We also examined traces of repackaged malware misclassified as benign apps. We found that the majority of such misclassified traces exhibited benign behaviors. The trained classifier, thus, reasonably assigned them to the benign class. This observation is expected, given that the test apps were stimulated using the primitive tool {\fontfamily{lms}\selectfont Droidutan}, which is unlikely to execute the malicious behaviors within repackaged malware after one execution.
Nevertheless, upon submitting the hashes of the apps corresponding to the examined traces to VirusTotal \cite{virustotal2017}, we noticed that a noticeable number of the malicious apps classified as benign were also deemed as benign by the majority of antiviral software. For example, the repackaged apps \emph{com.sojunsoju.go.launcherex.theme.greenlantern} and \emph{com.gametowin.save\_monster\_Google} were labeled as potentially unwanted by only two out of more than 50 antiviral software, whereas a repackaged version of \emph{com.gccstudio.DiamondCaveLite} was labeled as benign by all antiviral software on VirusTotal. The latter observation raises the question about the accuracy of the labels in current Android repackaged malware datasets, as discussed earlier.

In absence of the ground truth necessary to verify the reliability of the achieved scores, and based on our previous discussion and observations, our answer to question \textbf{Q2} is that we believe that the scores achieved by the active learning classifiers are likely to be reliable.

The active learning process we adopt to train classifiers might lead to overfitting. In other words, the trained classifier would not be able to correctly classify a large number of test apps (i.e., apps not used during the training phase). Research question \textbf{Q3} focuses on this issue. Similar to the training phase, we noticed that some classifiers underperform in comparison to conventional training methods, while others can match and outperform them. For example, using hybrid features, Random Forest classifiers trained using active learning manage to outperform conventional training methods using static, dynamic, and hybrid features. Furthermore, it avoids bias towards a certain class by maintaining a good balance between the F1 and specificty scores.
Observing the performances of different active learning classifiers, we conclude that Random Forest classifiers are able to achieve the best F1 and specificty scores during, both, training and test phases specifically using hybrid features, which is the answer to \textbf{Q4}.

\section{Related Work}
\label{sec:related}
To the best of our knowledge, there are no research efforts that attempt to apply active learning to the problem of stimulating and detecting Android repackaged malware. However, there are indeed efforts that study repackaged malware/piggybacked apps and attempt to stimulate and detect it. Moreover, we managed to identify research efforts that apply active learning to detecting Android malware in general. Consequently, we divide our related work section into two sub sections that discuss those two dimensions.
\subsection{Repackaged Malware Analysis and Detection}
There are different efforts that attempt to detect repackaging in Android apps including \cite{shao2014towards,hanna2012juxtapp,zhou2012detecting,deshotels2014droidlegacy}. Repackaged apps need not necessarily be malicious, however. An app could be repackaged for the sole purpose of injecting advertisement modules or translation purposes. In this paper, we focus on detecting repackaged malware, or piggybacked apps, that conform with Li et al.'s definition: benign apps that has been grafted with malicious payloads that are possible protected with trigger conditions \cite{li2017understanding}.

In \cite{pan2017dark}, Pan et al.\ developed a program analysis technique that decides whether an app witholds hidden sensitive operations (HSO) (i.e., malicious payloads), effectively deeming it as repackaged malware. Their technique, called \emph{HSOMiner}, gathers static features from the app's source code that capture the code segments' relationship with their surrounding segments, the type of inputs they tend to check, and the operations they usually perform. The extracted features are used to train and validate a SVM that decides whether an app contains a HSO. Needless to say, if an app contains HSO, it probably is malicious.
Tian et al.\ implemented another static approach to detect Android repackaged malware based on \emph{code heterogeneity analysis} \cite{tian2016analysis}. The approach is based on extract features that depict the dependence of a code segments on one another. In essence, injected code segments should exhibit more heterogeneity to the rest of the segments. That is to say, the malicious payloads injected into benign apps should be logically and semantically independent of the original app portions. Similar to our approach, they used the extracted features to train and compare the performance of four different classifiers viz., K-Nearest Neighbors, Support Vector Machine, Decision Tree, and Random Forest.
Lastly, Shahriar et al.\ use a metric called Kullback-Leibler Divergence (KLD) to detect Android repackaged malware \cite{shahriar2015kullback}. KLD is a metric depicting the difference between two probability distributions. In their paper, Shahriar et al.\ analyzed the Smali code of benign and malicious apps, from the Malgenome dataset, to build probability distributions of different Smali instructions. Their detection approach is based on the assumption that repackaged malware would have different distributions for different instructions than its benign counterpart.
\subsection{Detection with Active Learning}
Active learning has been applied to the problem of malware detection on the Windows operating system \cite{nissim2014novel} and within the network intrusion detection \cite{stokes2008aladin} domain. In \cite{zhao2012robotdroid}, Zhao et al.\ attempt to apply the same technique to the problem of Android malware detection. Despite adopting a similar architecture that comprises a characteristic (monitoring and extraction) module and a detection module, their work differs from ours in two main aspects. Firstly, Zhao et al.\ do not consider repackaged malware, rather Android malware in general. Secondly, and more importantly, their definition and utilization of active learning is different from ours. In \cite{zhao2012robotdroid}, active learning is defined as a technique to reduce the size of the training samples used to train a SVM classifier that deem an Android app, represented by a behavioral signature (i.e., API call trace), malicious or benign. This reduction is carried out as follows. A pool of pre-generated and labeled behavioral signatures is built as a source for training data. Given an unlabeled, out-of-sample signature, a classifier is trained using the subset of the training pool that guarantees maximum classification accuracy. The main difference between this method and ours is that the classification accuracy in \cite{zhao2012robotdroid} solely depends on the combination of the training samples, whereas ours depends on the content of such samples, especially since the training vectors usually differ from one iteration to another. Furthermore, this approach to active learning is very likely to yield small training datasets that hinder the trained classifiers from achieving good classification accuracies on test datasets (i.e., \emph{generalization}).
\section{Conclusion}
\label{sec:conclusion}
The lack of ground truth about the nature and location of malicious payloads injected into benign apps continues to hinder the development of effective methods to stimulate, analyze, and detect Android repackaged malware. We argue that training an effective classifier can be achieved if the classifier is allowed to request better representations of the training apps (i.e., via active learning).

Using a sample implementation of our proposed active learning architecture {\fontfamily{cms}\selectfont Aion}, we used active learning to stimulate, analyze, and detect Android repackaged malware. In \cite{zhou2012detecting}, Zhou et al.\ managed to detect malware in the Malgenome dataset with a best case of 79.6\% accuracy \cite{zhou2012detecting}, highlighting the difficulty of detecting such breed of malware. In this paper, our active learning classifiers managed to achieve test F1 scores of 72\% using the more recent dataset Piggybacking \cite{li2017understanding}, which we consider a benchmark for future comparison.

We consider this effort as the first step towards designing better app stimulation, analysis, and detection techniques that focus on Android repackaged malware, which continues to pose a serious threat \cite{wei2017deep}. Our future plan to enhance {\fontfamily{cms}\selectfont Aion} has three dimensions. Firstly, we plan on using the number of iterations it takes to train a classifier, and verify whether it transfers to the number of executions needed to effectively classify a test app. Secondly, we plan to use more sophisticated stimulation engines and compare their effect on the training and test scores compared to the random, primitive technique used in this paper. Thirdly, we plan to utilize features and classifiers that capture the semantics of the runtime behaviors exhibited by under test.

{\footnotesize \bibliographystyle{acm}
\bibliography{aion}}

\begin{thebibliography}{10}

\bibitem{apktool2017}
Apktool: https://goo.gl/eaewql.

\bibitem{droidutan2017}
Droidutan: https://goo.gl/7xsczq.

\bibitem{virustotal2017}
Virustotal: https://www.virustotal.com/en/search.

\bibitem{abraham2015}
{\sc Abraham, A., Andriatsimandefitra, R., Brunelat, A., Lalande, J.-F., and
  Viet Triem~Tong, V.}
\newblock {GroddDroid: a Gorilla for Triggering Malicious Behaviors}.
\newblock In {\em {10th International Conference on Malicious and Unwanted
  Software}\/} (Fajardo, Puerto Rico, 2015), {IEEE Computer Society}.

\bibitem{arp2014drebin}
{\sc Arp, D., Spreitzenbarth, M., Hubner, M., Gascon, H., and Rieck, K.}
\newblock Drebin: Effective and explainable detection of android malware in
  your pocket.
\newblock In {\em NDSS\/} (2014).

\bibitem{arshad2016android}
{\sc Arshad, S., Shah, M.~A., Khan, A., and Ahmed, M.}
\newblock Android malware detection \& protection: a survey.
\newblock {\em International Journal of Advanced Computer Science and
  Applications 7\/} (2016), 463--475.

\bibitem{bartel2012improving}
{\sc Bartel, A., Klein, J., Monperrus, M., Allix, K., and Le~Traon, Y.}
\newblock Improving privacy on android smartphones through in-vivo bytecode
  instrumentation.

\bibitem{deshotels2014droidlegacy}
{\sc Deshotels, L., Notani, V., and Lakhotia, A.}
\newblock Droidlegacy: Automated familial classification of android malware.
\newblock In {\em Proceedings of ACM SIGPLAN on Program Protection and Reverse
  Engineering Workshop 2014\/} (2014), ACM, p.~3.

\bibitem{fratantonio2016triggerscope}
{\sc Fratantonio, Y., Bianchi, A., Robertson, W., Kirda, E., Kruegel, C., and
  Vigna, G.}
\newblock Triggerscope: Towards detecting logic bombs in android applications.
\newblock In {\em Security and Privacy (SP), 2016 IEEE Symposium on\/} (2016),
  IEEE, pp.~377--396.

\bibitem{hanna2012juxtapp}
{\sc Hanna, S., Huang, L., Wu, E., Li, S., Chen, C., and Song, D.}
\newblock Juxtapp: A scalable system for detecting code reuse among android
  applications.
\newblock In {\em International Conference on Detection of Intrusions and
  Malware, and Vulnerability Assessment\/} (2012), Springer, pp.~62--81.

\bibitem{huang2013performance}
{\sc Huang, C.-Y., Tsai, Y.-T., and Hsu, C.-H.}
\newblock Performance evaluation on permission-based detection for android
  malware.
\newblock In {\em Advances in Intelligent Systems and Applications-Volume 2}.
  Springer, 2013, pp.~111--120.

\bibitem{li2017static}
{\sc Li, L., Bissyand{\'e}, T.~F., Papadakis, M., Rasthofer, S., Bartel, A.,
  Octeau, D., Klein, J., and Le~Traon, Y.}
\newblock Static analysis of android apps: A systematic literature review.
\newblock {\em Information and Software Technology\/} (2017).

\bibitem{li2017understanding}
{\sc Li, L., Li, D., Bissyand{\'e}, T.~F., Klein, J., Le~Traon, Y., Lo, D., and
  Cavallaro, L.}
\newblock Understanding android app piggybacking: A systematic study of
  malicious code grafting.
\newblock {\em IEEE Transactions on Information Forensics and Security 12}, 6
  (2017), 1269--1284.

\bibitem{li2017automatically}
{\sc Li, L., Li, D., Bissyande, T. F. D.~A., Klein, J., Cai, H., Lo, D., and
  Le~Traon, Y.}
\newblock Automatically locating malicious packages in piggybacked android
  apps.
\newblock In {\em 4th IEEE/ACM International Conference on Mobile Software
  Engineering and Systems\/} (2017).

\bibitem{luo+2014}
{\sc Luo, S., and Yan, P.}
\newblock Fake apps: Feigning legitimacy, 2014.

\bibitem{nissim2014novel}
{\sc Nissim, N., Moskovitch, R., Rokach, L., and Elovici, Y.}
\newblock Novel active learning methods for enhanced pc malware detection in
  windows os.
\newblock {\em Expert Systems with Applications 41\/} (2014), 5843--5857.

\bibitem{pan2017dark}
{\sc Pan, X., Wang, X., Duan, Y., Wang, X., and Yin, H.}
\newblock Dark hazard: Learning-based, large-scale discovery of hidden
  sensitive operations in android apps.

\bibitem{rasthofer2017making}
{\sc Rasthofer, S., Arzt, S., Triller, S., and Pradel, M.}
\newblock Making malory behave maliciously: Targeted fuzzing of android
  execution environments.
\newblock In {\em Proceedings of the 39th International Conference on Software
  Engineering\/} (2017), IEEE Press, pp.~300--311.

\bibitem{rasthofer2015current}
{\sc Rasthofer, S., Asrar, I., Huber, S., and Bodden, E.}
\newblock How current android malware seeks to evade automated code analysis.
\newblock In {\em IFIP International Conference on Information Security Theory
  and Practice\/} (2015), Springer, pp.~187--202.

\bibitem{sanz2013puma}
{\sc Sanz, B., Santos, I., Laorden, C., Ugarte-Pedrero, X., Bringas, P.~G., and
  {\'A}lvarez, G.}
\newblock Puma: Permission usage to detect malware in android.
\newblock In {\em International Joint Conference CISIS’12-ICEUTE{\'{}}
  12-SOCO{\'{}} 12 Special Sessions\/} (2013), Springer, pp.~289--298.

\bibitem{sato2013detecting}
{\sc Sato, R., Chiba, D., and Goto, S.}
\newblock Detecting android malware by analyzing manifest files.
\newblock {\em Proceedings of the Asia-Pacific Advanced Network 36\/} (2013),
  23--31.

\bibitem{shahriar2015kullback}
{\sc Shahriar, H., and Clincy, V.}
\newblock Kullback-leibler divergence based detection of repackaged android
  malware.

\bibitem{shao2014towards}
{\sc Shao, Y., Luo, X., Qian, C., Zhu, P., and Zhang, L.}
\newblock Towards a scalable resource-driven approach for detecting repackaged
  android applications.
\newblock In {\em Proceedings of the 30th Annual Computer Security Applications
  Conference\/} (2014), ACM, pp.~56--65.

\bibitem{stokes2008aladin}
{\sc Stokes, J.~W., Platt, J.~C., Kravis, J., and Shilman, M.}
\newblock Aladin: Active learning of anomalies to detect intrusion, 2008.

\bibitem{rednaga2016}
{\sc Strazzere, T.}
\newblock Detecting pirated and malicious android apps with apkid, 2016.

\bibitem{tam2017evolution}
{\sc Tam, K., Feizollah, A., Anuar, N.~B., Salleh, R., and Cavallaro, L.}
\newblock The evolution of android malware and android analysis techniques.
\newblock {\em ACM Computing Surveys (CSUR) 49}, 4 (2017), 76.

\bibitem{tian2016analysis}
{\sc Tian, K., Yao, D., Ryder, B.~G., and Tan, G.}
\newblock Analysis of code heterogeneity for high-precision classification of
  repackaged malware.
\newblock In {\em Security and Privacy Workshops (SPW), 2016 IEEE\/} (2016),
  IEEE, pp.~262--271.

\bibitem{tong2001active}
{\sc Tong, S.}
\newblock {\em Active learning: theory and applications}.
\newblock Stanford University, 2001.

\bibitem{wei2017deep}
{\sc Wei, F., Li, Y., Roy, S., Ou, X., and Zhou, W.}
\newblock Deep ground truth analysis of current android malware.
\newblock In {\em International Conference on Detection of Intrusions and
  Malware, and Vulnerability Assessment\/} (2017), Springer, pp.~252--276.

\bibitem{wu2012droidmat}
{\sc Wu, D.-J., Mao, C.-H., Wei, T.-E., Lee, H.-M., and Wu, K.-P.}
\newblock Droidmat: Android malware detection through manifest and api calls
  tracing.
\newblock In {\em Information Security (Asia JCIS), 2012 Seventh Asia Joint
  Conference on\/} (2012), IEEE, pp.~62--69.

\bibitem{xue2017malton}
{\sc Xue, L., Zhou, Y., Chen, T., Luo, X., and Gu, G.}
\newblock Malton: Towards on-device non-invasive mobile malware analysis for
  art.

\bibitem{zhao2012robotdroid}
{\sc Zhao, M., Zhang, T., Ge, F., and Yuan, Z.}
\newblock Robotdroid: A lightweight malware detection framework on smartphones.
\newblock {\em Journal of Networks 7}, 4 (2012), 715--722.

\bibitem{zhauniarovich2014}
{\sc Zhauniarovich, Y., Ahmad, M., Gadyatskaya, O., Crispo, B., and Massacci,
  F.}
\newblock {StaDynA: Addressing the Problem of Dynamic Code Updates in the
  Security Analysis of Android Applications}.
\newblock In {\em Proceedings of the 5th ACM Conference on Data and Application
  Security and Privacy\/} (2015), CODASPY '15, ACM, pp.~37--48.

\bibitem{zhou2012detecting}
{\sc Zhou, W., Zhou, Y., Jiang, X., and Ning, P.}
\newblock Detecting repackaged smartphone applications in third-party android
  marketplaces.
\newblock In {\em Proceedings of the second ACM conference on Data and
  Application Security and Privacy\/} (2012), ACM, pp.~317--326.

\bibitem{zhou2012dissecting}
{\sc Zhou, Y., and Jiang, X.}
\newblock Dissecting android malware: Characterization and evolution.
\newblock In {\em Security and Privacy (SP), 2012 IEEE Symposium on\/} (2012),
  IEEE, pp.~95--109.

\end{thebibliography}

\newpage
\section*{Appendix: Features}
\renewcommand{\thesubsection}{\Alph{subsection}}
\subsection{A: Static Features}
\label{app:appendixa}
In this section, we list the categories of static features extracted from the apps' APK's, and utilized in, both, preliminary and active learning experiments

\begin{table}[h]
\centering
\caption{The static features used during the preliminary static experiments.}
\label{tab:static_features}
\begin{tabular}{|c|c|}
\hline
Category & Feature \\ \hline
\multirow{6}{*}{basic} & Min. SDK version \\ \cline{2-2} 
 & Max. SDK version \\ \cline{2-2} 
 & Total \# of activities \\ \cline{2-2} 
 & Total \# of services \\ \cline{2-2} 
 & \begin{tabular}[c]{@{}c@{}}Total \# of broadcast\\ receivers\end{tabular} \\ \cline{2-2} 
 & \begin{tabular}[c]{@{}c@{}}Total \# of content\\ providers\end{tabular} \\ \hline
\multirow{4}{*}{permission} & \begin{tabular}[c]{@{}c@{}}Total requested\\ permissions\end{tabular} \\ \cline{2-2} 
 & \begin{tabular}[c]{@{}c@{}}Android permissions\\ $\div$\\ Total permissions\end{tabular} \\ \cline{2-2} 
 & \begin{tabular}[c]{@{}c@{}}Custom permissions\\ $\div$\\ Total permissions\end{tabular} \\ \cline{2-2} 
 & \begin{tabular}[c]{@{}c@{}}Dangerous permissions\\ $\div$\\ Total permissions\end{tabular} \\ \hline
API & \begin{tabular}[c]{@{}c@{}}Counts of API categories\\ listed in \href{https://aionexperiments.github.io/results/sensitiveAPICalls.py}{here}. \\ (Total: 27)\end{tabular} \\ \hline
\end{tabular}
\end{table}

\subsection{B: Dynamic Features}
\label{app:appendixb}
The following table lists the API categories used to extract dynamic features. The total number of API methods hooked is 71 methods, which makes them difficult to be listed in this paper. The exact list of API methods we hook can be found \href{https://aionexperiments.github.io/results/hooks.json}{here}.

\begin{table}[h]
\centering
\begin{small}
\begin{tabular}{|c|c|}
\hline
API Category                            & Total Hooked \\ \hline
android.accounts.AccountManager         & 2                    \\ \hline
android.app.Activity                    & 1                    \\ \hline
android.app.ActivityManager				& 1					\\ \hline
android.app.ActivityThread				& 1						\\ \hline
android.app.ApplicationPackageManager	& 2						\\ \hline
android.app.ContextImpl					& 1						\\ \hline
android.app.NotificationManager         & 1                    \\ \hline
android.app.SharedPreferencesImpl\$EditorImpl & 5					\\ \hline
android.content.BroadcastReceiver		&	1					\\ \hline
android.content.ContentResolver         & 4                     \\ \hline
android.content.ContentValues           & 1                   \\ \hline
android.location.Location				& 2						\\ \hline
android.media.AudioRecord				& 1						\\ \hline
android.media.MediaRecorder				& 1						\\ \hline
android.net.ConnectivityManager		    & 1						\\ \hline
android.net.wifi.WifiInfo				& 1						\\ \hline
android.os.Debug						& 1						\\ \hline
android.os.Process						& 1						\\ \hline
android.os.SystemProperties				& 1						\\ \hline
android.telephony.SmsManager			& 2						\\ \hline
android.telephony.TelephonyManager		& 11						\\ \hline
android.util.Base64						& 3						\\ \hline
android.system.BaseDexClassLoader		& 3						\\ \hline
android.system.DexClassLoader			& 3						\\ \hline
android.system.DexFile					& 2 						\\ \hline
android.system.PathClassLoader			& 3						\\ \hline
java.io.FileInputStream					& 1						\\ \hline
java.io.FileOutputStream				& 1						\\ \hline
java.lang.ProcessBuilder				& 1						\\ \hline
java.lang.Runtime						& 1						\\ \hline
java.lang.reflect.Method				& 1						\\ \hline
java.net.URL 							& 1						\\ \hline
javax.crypto.Cipher						& 1						\\ \hline
javax.crypto.Mac						& 1						\\ \hline
javax.crypto.spec.SecretKeySpec			& 5						\\ \hline
libcore.io.IoBridge 					& 1						\\ \hline
org.apache.http.impl.client.AbstractHttpClient				& 1					\\ \hline
\end{tabular}
\end{small}
\caption{The API categories that comprise the dynamic features used in the preliminary dynamic and active learning experiments.}
\label{tab:dynamic_features}
\end{table}

\subsection{C: Figures}
\label{app:appendixc}

This section contains the plots of specificity scores achieved by various classifiers during the static and dynamic variations of the preliminary experiments.
\begin{figure*}
\centering
\begin{subfigure}{0.49\textwidth}
	\scalebox{.55}{\input{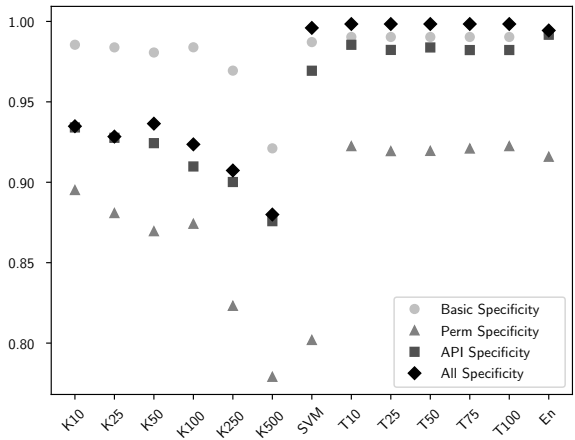}}
	\caption{Malgenome median specificity}
	\label{fig:exp_pre_static_malgenomesp}
\end{subfigure}
\begin{subfigure}{0.49\textwidth}
	\scalebox{.55}{\input{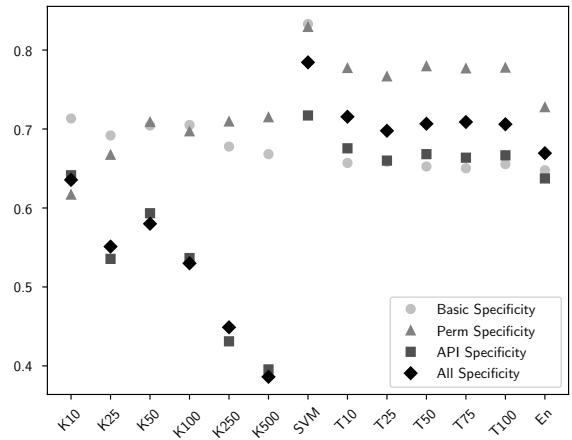}}
	\caption{Piggybacking median specificity}
	\label{fig:exp_pre_static_piggybackingsp}
\end{subfigure}
\caption{The median (after 25 runs) specificity scores (Y-axis) recorded by 13 classifiers (X-axis) using static features on the test datasets.}
\label{fig:exp_pre_static_sp}
\end{figure*}

\begin{figure}
\centering
\scalebox{.55}{\input{figs/exp_preliminary_dynamic/Piggybacking_dynamic_TEST_Specificity_scatter.pgf}}
\caption{The median (after 25 runs) specificity scores (Y-axis) recorded by 13 classifiers (X-axis) using dynamic and hybrid features versus the best scoring static features (permission-based) on the test datasets of Piggybacking.}
\label{fig:exp_pre_dynamic_sp}
\end{figure}


\end{document}